\documentclass[sn-chicago]{sn-jnl}

\usepackage[rightcaption]{sidecap}

\begin{document}

\title[Locating $\gamma$-Ray Sources via Modal Clustering]{Locating $\gamma$-Ray Sources on the Celestial Sphere via Modal Clustering}

\author[1]{\fnm{Anna} \sur{Montin}}\email{anna.montin@studenti.unipd.it}
\equalcont{These authors contributed equally to this work.}

\author*[1]{\fnm{Alessandra R.} \sur{Brazzale}}\email{alessandra.brazzale@unipd.it}
\equalcont{These authors contributed equally to this work.}

\author[1]{\fnm{Giovanna} \sur{Menardi}}\email{giovanna.menardi@unipd.it}
\equalcont{These authors contributed equally to this work.}

\affil*[1]{\orgdiv{Department of Statistical Sciences}, \orgname{University of Padova}, \orgaddress{\street{Via C. Battisti 241}, \city{Padova}, \postcode{35121}, \state{(PD)}, \country{Italy}}}

\abstract{%
Searching for as yet undetected $\gamma$-ray sources is a major target of the Fermi LAT Collaboration.  We present an algorithm capable of identifying such type of sources by non-parametrically clustering the directions of arrival of the high-energy photons detected by the telescope onboard the Fermi spacecraft.  In particular, the sources will be identified using a von Mises-Fisher kernel estimate of the photon count density on the unit sphere via an adjustment of the mean-shift algorithm to account for the directional nature of data.  This choice entails a number of desirable benefits.  It allows us to by-pass the difficulties inherent on the borders of any projection of the photon directions onto a 2-dimensional plane, while guaranteeing high flexibility.  The smoothing parameter will be chosen adaptively, by combining scientific input with optimal selection guidelines, as known from the literature.  Using statistical tools from hypothesis testing and classification, we furthermore present an automatic way to skim off sound candidate sources from the $\gamma$-ray emitting diffuse background and to quantify their significance.  The algorithm was calibrated on simulated data provided by the Fermi LAT Collaboration and will be illustrated on a real Fermi LAT case-study.%
}


\maketitle

\section{Motivation and rationale}
\label{sec:motivation-and-background}

\subsection{High-energy astrophysics}
\label{high-energy-astrophysics}
The past three decades have been a golden era for Astronomy.  Pioneering technology has driven remarkable acceleration in the rate of detection and characterization of celestial objects, and new space missions will have more and better quality data to help find and characterize these objects.  Discoveries in this field are of utmost relevance as they contain a wealth of information about the history of the Universe, and impact on the understanding of our Galaxy and our own Solar system.  An important example is high-energy astrophysics, which acts at the interface between particle physics and astronomy to study the multitude of extreme phenomena which inhabit the Cosmos.  To date, the observation of $\gamma$-ray photons, that is, of quanta of light in the highest energy range, has provided the basis for a large number of astronomical discoveries.  $\gamma$-rays are usually generated from accelerated charged particles, such as electrons or protons, boosted by extreme celestial objects such as supermassive black holes, supernova remnants, pulsars and active galactic nuclei, to name a few.  The study of these $\gamma$-ray emitting sources improves our understanding of high-energy astrophysical phenomena, and might even resolve the mystery of the fundamental nature of dark matter.

The Fermi Gamma-ray Space Telescope is an international and multi-agency space mission launched in June 2008 which studies the Cosmos in the energy range 10 keV -- 300 GeV.  The primary instrument onboard the Fermi spacecraft is the Large Area Telescope (LAT), a wide field-of-view pair-conversion telescope which was designed to perform an all-sky survey aimed at discovering and locating high-energy emitting sources.  The standard procedure of the Fermi LAT Collaboration for point-like source detection relies on so-called {\it single-source} models \cite[par.~7.4]{Hobson-etal-2009}, which require the sky map to be split into small regions.  The presence of a possible new source is assessed on a pixel-by-pixel basis: Poisson regression is used to model the number of photons associated with each pixel and likelihood ratio tests assess the significance of the source \citep{MattoxJ}.  See also \cite{van_Dyk_2001} for a Bayesian treatment with appplication to low-count X-ray data collected by the Chandra X-Ray Observatory.  Conversely, {\it variable-source-number} models address the problem from a more global perspective, as they simultaneously identify and locate all possible sources in a given sky map \cite[par.~7.3]{Hobson-etal-2009}.  
Since point-like sources present themselves as spatially concentrated photon emissions, the problem can naturally be recast as a clustering problem.  Recent examples of variable-source-number modelling of X-ray and $\gamma$-ray photon count data using finite and infinite mixtures are 
\cite{Jones2015}, \cite{Costantin2020primo}, \cite{Costantin2020due}, \cite{Sottosanti2021} and \cite{Meyer2021}.  

\begin{figure}[t]
\begin{center}
\includegraphics[height=0.15\textheight]{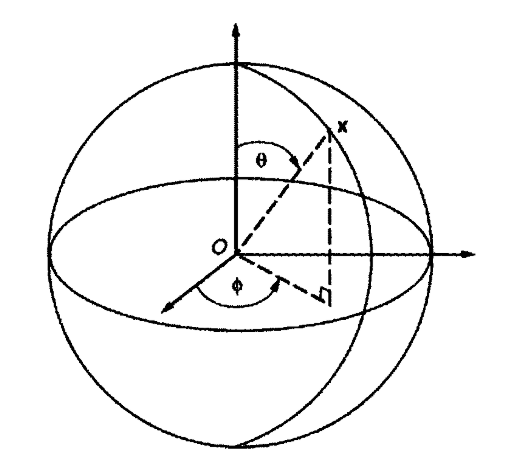}  \,
\includegraphics[height=0.14\textheight]{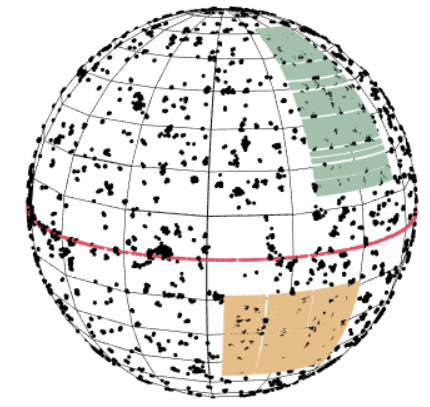} \,
\includegraphics[height=0.14\textheight]{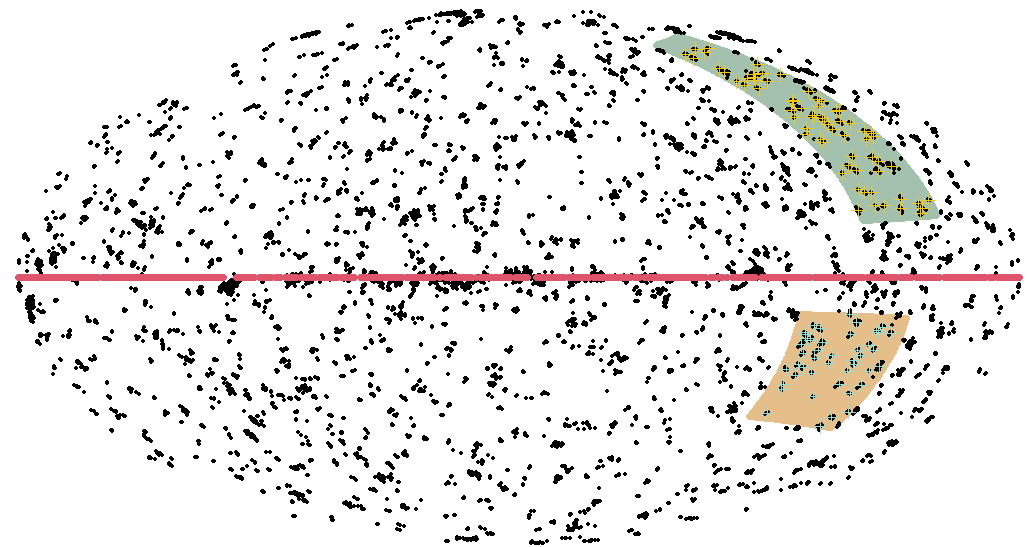} 
\end{center}
\caption{Left: Polar coordinates as recorded by the LAT \citep[Image credit:][]{Mardia2000}.  $\phi$ is the longitude and $\theta$ is the co-latitude.  Center and right: Fermi-LAT $\gamma$-ray photon count maps for a \mbox{5-year} observation period.  Center: in polar coordinates.  Right: in Galactic coordinates.  Yellow: region of size $(l,b) \in [95^\circ,135^\circ] \times [-40^\circ,-10^\circ]$ analyzed in Section~\ref{case-study}.  Green: photon counts used to train the post-processing classifier.  Red: Galactic plane.}
\label{fig:1}       
\end{figure}

The data provided by the Fermi LAT Collaboration typically consist of an event list which gives the direction in the sky of each detected photon together with additional information, the primary one being its energy content and the so-called event type which expresses the quality of the measurement.  This information is used to determine the number of the emitting extra-galactic sources, measure their intensities, and assign to them the corresponding individual photon counts.  A major challenge of trying and detecting high-energy phenomena from astronomical data is to separate the signal of the putative emitting source from noise.  The Fermi LAT data, in particular, are characterized by two types of noise: (i) measurement error associated with the components of the LAT (tracker, calorimeter etc.) and (ii) the diffuse $\gamma$-ray background which spreads over the entire area observed by the telescope.  The former is expressed through the LAT's {\it point spread function} \citep{Ackermann2013}, which is typically included into the model.  Different phenomena contribute to the residual $\gamma$-ray background \citep{Acero_2016}.  Broadly speaking, its origins can be brought under two headings: galactic interstellar emission (GIE), that is, the interaction of galactic cosmic rays with gas and radiation fields, and a residual all-sky emission.  The latter is commonly called the isotropic diffuse gamma-ray background (IGRB), and includes the $\gamma$-ray emission from faint unresolved sources and any residual galactic emission which is approximately isotropic.  \cite{Costantin2020due} translate the simulation-based background model developed by \cite{Acero_2016} into a workable parametric formulation, while  \cite{Sottosanti2021} reconstruct it via a flexible Bayesian nonparametric model based on B-splines.  

A further challenge of analysing the Fermi LAT data refers to the geometry of the problem.  As the distance to the emitting source is not given, the data points are placed on the celestial sphere with Earth at its center and unit radius, as shown in the middle panel of Figure~\ref{fig:1}.\nocite{Mardia2000}  Directions are expressed in \textit{Galactic coordinates}, that is longitude $l$ and latitude $b$, which place the origin of the Cartesian system in the center of our galaxy --- the Milky Way --- and align the $x$-axis with the Galactic plane (right panel of Figure~\ref{fig:1}).  This is the situation considered by \cite{Jones2015}, \cite{Costantin2020due}, \cite{Sottosanti2021} and \cite{Meyer2021}.  Instead of projecting data onto a 2-dimensional map, we may rather express directions in 3 dimensions through \textit{polar coordinates}, that is, co-latitude ($\theta$) and longitude ($\phi$) in geographical terms; see the left panel of Figure~\ref{fig:1}.  These can easily be back-transformed to Cartesian coordinates $\mathbf{x} = [\cos\theta, \sin\theta\cos\phi, \sin\theta\sin\phi]^\top$ on the unit sphere, as done by \cite{Costantin2020primo}.  A thorough treatment of directional data can be found in \cite{Mardia2000}.

\subsection{The statistical state of the art}
The discovery of celestial objects is an intrinsically interdisciplinary field which combines both, statistical and astrophysical methodology.  Statistical learning, by which we mean the ability of discovering patterns and regularities in the data, plays a central role in knowledge discovery.  This also includes allocating objects to a pre-assigned or unknown number of groups according to a set of observed attributes or features, which is a natural activity of any science.  A major distinction is made depending on whether the groups are defined, and known a priori, or need be detected using the data.  Clustering, or unsupervised learning, considers the latter situation.  A surge of techniques has been proposed over the years, which differ significantly in their definition of what a cluster is and how to identify it \citep{Hennig-etal-2015}.  A precise statistical notion of what a ``group'' is, is provided by the {\it density-based} approach.  Here, the clusters are associated with some specific features of the probability distribution which is assumed to underlie the data.  This idea has been developed into two distinct directions. The {\it model-based} or parametric approach represents the probability distribution of the data as a mixture of parametric distributions.  A cluster is associated with each component of the mixture and the observations are allocated to the cluster with maximal density among the components.  Standard accounts are the seminal works of \cite{FraleyRaftery1998,FraleyRaftery2002}.  A less widespread density-based clustering formulation is referred to as {\it modal} or nonparametric clustering and dates back to \cite{Carmichael1968}.  Here, the underlying density is reconstructed from the data using suitable nonparametric density estimators, and clusters are associated with the domain of attraction of their modes. The rather scattered theory is reviewed in \cite{Menardi2016}.  \cite{Chacon2015} provides some new insight into the theoretical foundations of modal clustering making
use of Morse theory \citep{Milnor1969}.   

In this paper, we advocate the use of nonparametric, or modal clustering for $\gamma$-ray source detection using a von Mises-Fisher kernel on the unit sphere.  This choice entails a number of desirable benefits.  It allows us to by-pass the difficulties inherent on the borders of any 2-dimensional projection of the photon directions.  But, it also guarantees high flexibility and adaptability, while posing on a sound theoretical ground.  The sources will be identified via an adjustment of the mean-shift algorithm to account for the directional nature of the Fermi LAT data.  The issue of selecting the smoothing parameter is addressed adaptively, by combining scientific input with optimal selection guidelines, as known from the literature.  Using known results from hypothesis testing and classification, we furthermore present an automatic way to pinpoint sound candidate sources and to quantify their significance by skimming off the $\gamma$-ray emitting diffuse background.  The Fermi LAT database currently holds over 1 billion photons in the energy range from about 20 MeV to more than 300 GeV collected in over a decade of operation.  Efficient tools to account for the computational burden required to analyse huge amounts of data, possibly on the entire sphere, are also discussed.  Our method was calibrated on simulated data provided by the Fermi LAT Collaboration and will be illustrated on a real Fermi LAT case-study.

The paper is organized as follows.  Section~\ref{kernel-density-estimator-for-directional-data} sets the methodological background of kernel density estimation for directional data.  Being able to correctly specify the right amount of smoothing is crucial for the reliable identification of the sources.  Optimal bandwidth selection is discussed in Section~\ref{bandwidth-selection}, while Section~\ref{modal-clustering-on-the-unit-sphere} presents our proposal of modal clustering on the unit sphere.  In particular, to separate the true signal emitted by a source from the background, we developed a post-processing procedure that combines the findings of two parallel quests.  One establishes the significance of a candidate mode using a suitable statistical test as presented in Section~\ref{statistical-significance}.  The second skims off the photons emitted by the $\gamma$-ray background using a tree-based classifier build on previous knowledge provided by the Fermi LAT Collaboration; see Section~\ref{feature-selection}.  Section~\ref{benchmarking} benchmarks two key aspects of our proposal, namely the selection of the optimal bandwidth and the classification of the incoming photons.  Section~\ref{case-study} eventually illustrates the performance of our proposal when applied to a real sample of high-energy photons accumulated by the LAT.  The paper closes with the concluding remarks of Section~\ref{conclusions}.

This paper is an extended version of the paper presented at the 51st Scientific Meeting of the Italian Statistical Society on June, 2022 \citep{Montin2022}.

\section{Kernel density estimators for directional data}
\label{kernel-density-estimator-for-directional-data}

\subsection{The von Mises-Fisher distribution}
Directions in the 3-dimensional space can be represented using Cartesian coordinates as unit vectors $\mathbf{x}$, that is, as points on the sphere 
$$\Omega_ 2 = \{ \boldsymbol{x} \in \mathbb{R}^{3}: \|\boldsymbol{x}\|_2 = x_1^2 + x_2^2 + x_3^2 = 1\}$$ 
with unit radius and centre at the origin.  These can be retrieved from Galactic coordinates, that is, from the longitude $l\in(-180,+180)$ and the latitude $b\in(-90,+90)$ of a given data point, by 
$$\mathbf{x} = [\cos l\cos b, \ \sin l\cos b,\ \sin b]^\top.$$
A widely used distribution to model $\gamma$-ray emission in astrophysics searches \citep{Banerjee2006} is the von Mises-Fisher (vMF) distribution 
\begin{equation}
\nonumber
f_{vMF}(\boldsymbol{x}; \boldsymbol{\mu}, \kappa) = C_2(\kappa) \exp\{ \kappa \boldsymbol{x}^\top\boldsymbol{\mu}\},
\end{equation}
which extends the 3-dimensional normal distribution $N_3(\boldsymbol{\mu}, \kappa^{-1} \text{\bf I}_3)$, with $\text{\bf I}_3$ being the $3\times3$ diagonal unit matrix, by restricting its density to the unit sphere.  Here, $\boldsymbol{\mu} \in \Omega_2$ represents the mean direction, while $\kappa \geq 0$ is a concentration parameter \citep[Section~9.6]{Mardia2000}.  As such, the von Mises-Fisher distribution describes observations which scatter simmetrically around their mean direction $\boldsymbol{\mu}$.  The normalizing constant
\[
C_2(\kappa) = \frac{\kappa^{\frac{1}{2}}}{(2\pi)^{\frac{3}{2}}\mathcal{I}_{\frac{1}{2}}(\kappa)}
\]
includes the modified Bessel function 
\begin{equation}
\nonumber
\mathcal{I}_{\nu}(z) = \frac{\Big(\frac{z}{2}\Big)^{\nu}}{\pi^{1/2}\Gamma(\nu+\frac{1}{2})} \int_{-1}^1 (1-t^2)^{\nu-\frac{1}{2}} e^{zt} dt 
\end{equation}
of order $\nu=1/2$.

\subsection{Kernel density estimator}
Let $\mathbf{x}_1,\ldots,\mathbf{x}_n \in \Omega_2$ be a random sample of $n$ observations generated by a distribution with density $f(\mathbf{x})$ defined on the unit sphere $\Omega_2$ such that
\begin{equation}
\nonumber
\int_{\Omega_2}f(\mathbf{x})\omega_2(d\mathbf{x}) = 1,
\end{equation}
where $\omega_2$ is the Lebesgue measure on $\Omega_2$.  We can estimate the density $f$ using the kernel density estimator proposed by \cite{Bai1988} for directional data, 
\begin{equation}
\label{directional_kernel}
\hat f_h(\boldsymbol{x}) = \frac{c_{h}(K)}{n} \sum_{i=1}^n K \biggl( \frac{1-\boldsymbol{x}^\top \boldsymbol{x}_i}{h^2}\biggr),
\end{equation}
where $K(\cdot)$ is a suitable kernel function which decreases on $[0, \infty)$, and $h>0$ is the smoothing parameter.  The normalizing constant $c_{h}(K)$, is defined by
\[
c_{h}(K)^{-1} = \int_{\Omega_2} K \biggl( \frac{1-\boldsymbol{x}^\top \boldsymbol{x}_i}{h^2} \biggr) \omega_2(d\boldsymbol{x})  = h^2 \tilde c_{h}(K),
\]
where $\tilde c_{h}(K) = \int_0^{2/h^{2}} K(u)du$.  
Using the von Mises-Fisher kernel, expression (\ref{directional_kernel}) becomes
\begin{equation}
\label{vMF_directional_kernel}
\begin{split}
\hat f_h(\boldsymbol{x})
&  =  \frac{1}{n} \sum_{i=1}^n f_{vMF}\biggl( \boldsymbol{x}; \boldsymbol{x}_i, \frac{1}{h^2}\biggr) \\
& 
= \frac{1}{(2\pi)^{\frac{3}{2}} \mathcal{I}_{\frac{1}{2}}(h^{-2})} \frac{1}{hn} \sum_{i=1}^n \exp\biggl( \frac{\boldsymbol{x}^\top \boldsymbol{x}_i}{h^2}\biggr).
\end{split}
\end{equation}
That is, the kernel density estimator for direction data on the unit sphere is a mixture of 3-dimensional von Mises-Fisher distributions with $\kappa= h^{-2}$.

\begin{figure}[t]
\begin{center}
\includegraphics[height=0.16\textheight]{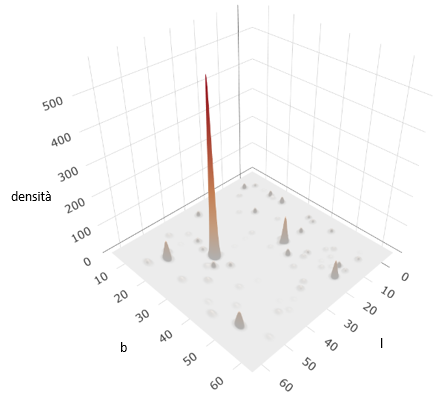} \,
\includegraphics[height=0.16\textheight]{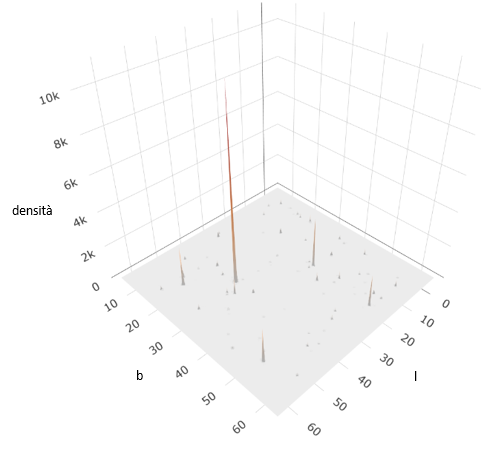} \,
\includegraphics[height=0.16\textheight]{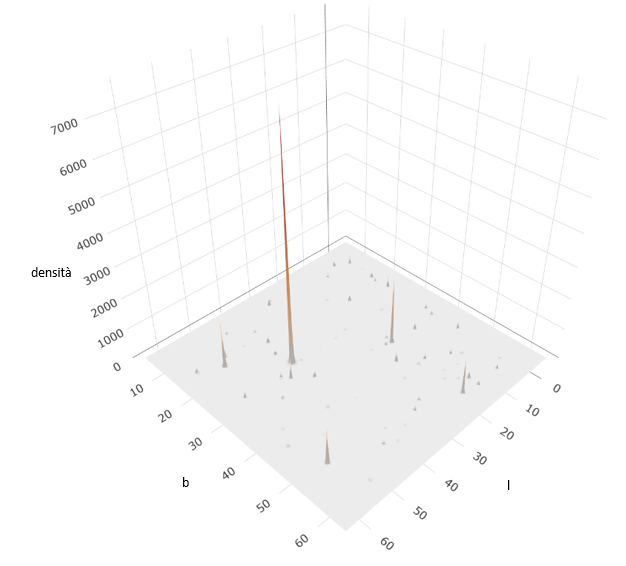} 
\end{center}
\caption{von Mises-Fisher kernel density estimate of the high-energy photons tracked by the Fermi LAT in the validation region $(l,b) \in [0^\circ,60^\circ] \times [10^\circ,60^\circ]$ for different values of $h$: 0.01 (left), 0.001 (center), $h_{i,SE}$ (right).}
\label{fig:4}       
\end{figure}

\section{Bandwidth selection} 
\label{bandwidth-selection}

\subsection{Data-based methods}
\label{data-based-methods}
A major issue when using a kernel density estimator is the selection of the smoothing parameter, or bandwidth, $h$.  Being able to correctly specify the right amount of smoothing is crucial for the reliable identification of the sources.  This is illustrated in Figure~\ref{fig:4}, which plots the estimated density for the same sky region using three different values of $h$, where the latter choice varies with sky location.  If the smoothing parameter is too large (picture on the left), false peaks may emerge from the background.  Conversely, if the kernel function is too concentrated (middle picture), we may miss some faint sources.  A wealth of data-driven methods were developed over the years for both, fixed and variable bandwidth kernel density estimation.  As far as directional data goes, the proposals mainly are for circular observations; see e.g.\ \cite{10.1093/biomet/74.4.751} and \cite{KLEMELA200018}.  Adaptive kernel density estimation, that is, when the smoothing parameter $h_i$ in \eqref{vMF_directional_kernel} adapts to the local behaviour of $f$ at $\boldsymbol{x}_i$, is of special interest to us, as the spatial scattering of the incoming photons differs among sources, and to an even larger extent if they were emitted from the background radiation.    

\begin{figure}[t]
\begin{center}
\includegraphics[height=0.28\textheight]{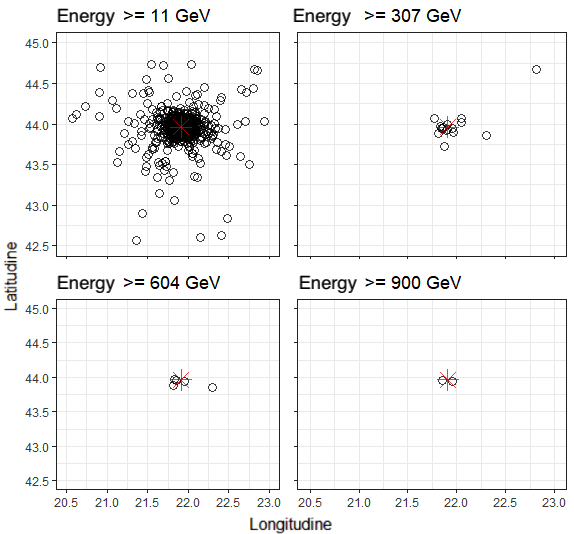} \,
\includegraphics[height=0.28\textheight]{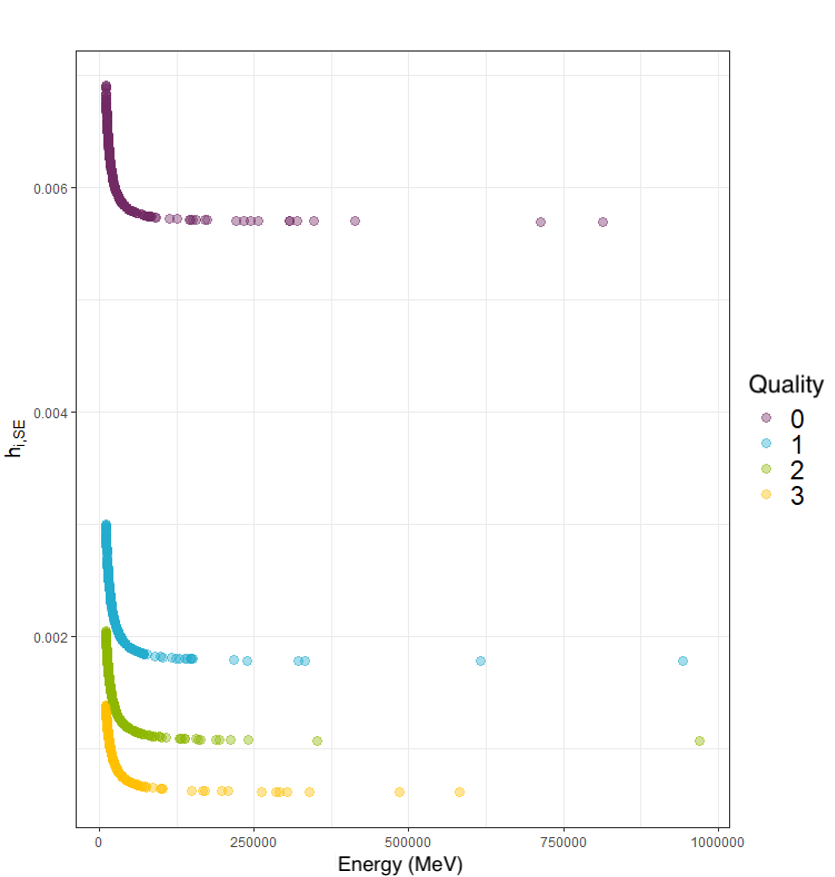}
\end{center}
\caption{Left: Photon scattering as a function of their energy content \citep[courtesy of][]{Sottosanti2021}.  Right: Values of $h_{i,SE}$ as a function of energy and event quality, where PSF0 represents the worst event type.  The higher the energy and quality of the event, the smaller is the smoothing parameter.}
\label{fig:2}       
\end{figure}

Selecting an optimal bandwidth generally entails minimization of a suitable measure of the error we commit when estimating the target density $f$ by $\hat f_h$.  A common way of measuring this error is the mean integrated squared error 
\[ 
MISE(h) = \mathbb{E}\left[ \int_{\Omega_2} \left\{\hat f_h(\boldsymbol{x}) - f(\boldsymbol{x})\right\}^2\omega_2(d\boldsymbol{x}) \right ],
\]
where the expectation is taken with respect to the distribution specified by $f$; in this case
\[ 
h_{MISE} =\arg\min_{h>0} MISE(h).
\]
The alternative choice $h_{AMISE}$ minimizes the asymptotic approximation of the mean integrated squared error, that is, when $n\rightarrow\infty$.  However, both window widths depend explicitly on the unknown density to be estimated, and cannot be computed exactly.  Simple ``plug-in" procedures, where $f$ is replaced by a suitable pilot estimate $\hat f$, turned out to be generally unsatisfactory.  An automatic way of determining the optimal bandwidth $h$ is by {\it likelihood cross-validation}, that is,
\[
h_{LCV} = \arg \max_{h>0}\ CV (h), \quad \text{where} \quad CV(h) = \sum_{i=1}^n \log \hat f_{h,-i}(\boldsymbol{x}_i).
\]
Here, $\hat f_{h,-i}(\boldsymbol{x}_i)$ is the kernel density estimate we obtain after having omitted observation $i$, evaluated at $\boldsymbol{x}_i$.  A further option is to adapt the most promising solutions for optimal bandwidth selection on the plane to our problem at hand, as listed below.  The corresponding performance metrics were evaluated on the simulated sample of high-energy photons emitted by the sources present in the sky region shown in Figure~\ref{fig:4}, and will be discussed in Section~\ref{optimal-bandwidth}.  

A first possibility is to generalize Garc\'\i a-Portugu\' es' (2013) \nocite{Garcia-Portugues2013} {\it rule of thumb} to spherical data, 
\[
h_{THUMB} = \Biggl\{ \frac{8\sinh^2(\hat \kappa)}{\hat \kappa [(1+4\hat \kappa^2) \sinh(2\hat \kappa)-2\hat \kappa \cosh(2\hat \kappa) ]n} \Biggr\}^{\frac{1}{6}},
\]
where the concentration parameter $\hat \kappa$ is estimated by maximum likelihood.  Conversely, if we want the bandwidth $h$ to depend on the current location $\boldsymbol{x}_i$ of the estimator, a first possibility is to use Abramson's (1982) \nocite{Abramson0a} rule, which has $h_i$ change proportionally with the inverse of the square root of $\hat f_{h}(\boldsymbol{x}_i)$, 
\[
h^A_{i,THUMB} = h_{THUMB} \bigg[ \hat f_{h_{THUMB}}(\boldsymbol{x}_i)\bigg]^{-\frac{1}{2}} \quad \text{and} \quad h^A_{i,LCV} = h_{LCV} \bigg[ \hat f_{h_{LCV}}(\boldsymbol{x}_i)\bigg]^{-\frac{1}{2}}.
\]
Here, $\hat f_{h_{THUMB}}(\boldsymbol{x}_i)$ and $\hat f_{h_{LCV}}(\boldsymbol{x}_i)$ are winsorized (or clipped) versions of a suitably constructed pilot kernel density estimate with fixed bandwidth $h$, which may be $h_{THUMB}$ or $h_{LCV}$.  A second possibility is to use the modification proposed by \citet[Section~5.3]{silverman2018density}, 
\[
h^{S}_{i, THUMB} = h_{THUMB} \bigg[ \frac{1}{m_g} \hat f_{h_{THUMB}}(\boldsymbol{x}_i)\bigg]^{-\beta} \quad \text{and} \quad h^{S}_{i, LCV} = h_{LCV} \bigg[ \frac{1}{m_g} \hat f_{h_{LVC}}(\boldsymbol{x}_i)\bigg]^{-\beta},
\]
where $m_g$ is a scale factor defined by the geometric mean of the two pilot estimates, $\hat f_{h_{THUMB}}(\boldsymbol{x}_i)$ or $\hat f_{h_{LCV}}(\boldsymbol{x}_i)$, while $\beta\in[0,1]$ tunes the sensitivity of the bandwidth to variations of these.  We will set $\beta=0.5$, as this choice generally entails a better behavior of the kernel density estimator on the tails of the distribution \citep{Izenman91}.

\subsection{Using scientific input}
\label{using-scientific-input}
A valid alternative for determining the smoothing parameter $h$ is to use scientific input.  As mentioned in Section~\ref{high-energy-astrophysics}, the spatial scattering of the photons around the source direction $\boldsymbol{\mu}$ is modelled by the LAT's \textit{point spread function} (PSF).  This function depends on the energy of the incoming photon, on its inclination angle $\theta$ (see left panel of Figure~\ref{fig:1}) and on the quality of the recorded event \citep{Ackermann2013}.  The latter is expressed by the PSF event type, that is, an event-level quantity which indicates how well the LAT managed to reconstruct the direction of the incoming photon and which assumes four values, from the lowest quality (PSF0) to the best quality (PSF3).  Most importantly, the PSF depends on the scale factor 
\begin{equation}
\nonumber
S(E_i) \propto \sqrt{\Big[ c_{0,i} \Big( \frac{E_i}{100\text{MeV}}\Big)^{-0.8} \Big]^2 + c_{1,i}^2},
\end{equation}
which describes the uncertainty of the event as a decreasing function of the energy $E_i$, expressed in Mega electron Volt (MeV), and of the two parameters $c_{0,i}$ and $c_{1,i}$, which are given distinct values for the different event qualities and can be retrieved from the Fermi LAT web site\footnote{https://fermi.gsfc.nasa.gov/ssc/data/analysis/documentation/Cicerone/Cicerone\_LAT\_IRFs/IRF\_PSF.html}.  The first constant, $c_{i,0}$, represents multiple scattering while the second, $c_{1,i}$, represents the spatial resolution of the LAT tracker.  How the precision of the measurements depends on the energy is shown in the left panel of Figure~\ref{fig:2} \citep{Sottosanti2021}, while the right panel of the same figure plots the values we obtain for $h_{i, SE}$ for the four different event types.  
On this basis, we may specify a variable bandwidth as
\begin{equation}
h_{i, SE} = \sqrt{\Big( c_{0,i} \Big( \frac{E_i}{100\text{MeV}}\Big)^{-0.8} \Big)^2 + c_{1,i}^2}, 
\end{equation}
which is the one used in the right panel of Figure~\ref{fig:4}.

\section{Modal clustering on the unit sphere}
\label{modal-clustering-on-the-unit-sphere}

\subsection{Mode hunting}
\label{sec:ms_dir}
Modal clustering associates clusters with the domain of attraction of the modes of the underlying density $f$.  Two main strands can be identified, depending on whether the modes are given explicitely or not \citep{Menardi2016}.  A first strand follows the route of \cite{Hartigan75} and identifies clusters with high-density regions of the sample space, defined by the density level sets
\[
L_c(f) = \{ \mathbf{x} \in \Omega_2 : f(\mathbf{x}) \geq c\}, \quad 0\leq c \leq \max f.
\]
An estimate of the unknown $L_c(f)$ is obtained by replacing $f(\mathbf{x})$ by its non-parametric estimate $\hat f(\mathbf{x})$.  The rationale behind this class of methods is that any connected component of $L_c(f)$ includes at least one mode of the density function, and, on the other hand, for each mode of the density function, there exists $\lambda$ for which one of the connected components of the associated $L(\lambda)$ includes this mode at most.  The major drawback is that the identification of the connected components of a multidimensional set is not straightforward. 

As our aim is to discover and identify unknown $\gamma$-ray emitting sources, we want to associate their direction explicitly with the modes of the unknown density $f$.  \cite{Yang2014} adapted the {\it mean-shift} algorithm developed by \cite{Fukunaga1975} to be used with the directional kernel estimator \eqref{vMF_directional_kernel} and fixed bandwidth $h$.  Starting from a generic point $\mathbf{x}^{(0)}$, the algorithm recursively shifts it to a local weighted mean, until convergence.  Denoted by $w_i(\mathbf{x}^{(s)})$ the vector of weights of the components of $\mathbf{x}_i$ at step $s$, at the next step, $(s+1)$, we have
\[
\mathbf{x}^{(s+1)} = \sum_{i=1}^n w_i(\mathbf{x}^{(s)})\mathbf{x}_i = \mathbf{x}^{(s)} + M(\mathbf{x}^{(s)}), 
\]
where $M(\mathbf{x}^{(s)}) = \sum_{i=1}^n w_i(\mathbf{x}^{(s)})\mathbf{x}_i - \mathbf{x}^{(s)}$ denotes the mean shift.  Up to a normalising factor, the weights $w_i(\mathbf{x})$ involve the derivative $K^\prime(h^{-2}(1-\mathbf{x}^\top \mathbf{x}_i))$ of the kernel function, which leads to the weighted average
\begin{equation}
\nonumber
\hat{\boldsymbol{x}}^{(s+1)} = - \frac{\sum_{i=1}^n \boldsymbol{x}_i K'\biggl(\frac{1-\hat{\boldsymbol{x}}^{(s)\top} \boldsymbol{x}_i}{h^2} \biggr)}{\Big\vert\Big\vert \sum_{i=1}^n \boldsymbol{x}_i K'\biggl( \frac{1-\hat{\boldsymbol{x}}^{(s)\top}\boldsymbol{x}_i}{h^2} \biggr)\Big\vert\Big\vert_2},
\end{equation}
where $\mid\mid \cdot \mid\mid_2$ is the Euclidean norm.  Here, the minus sign is due becasue $K(\cdot)$ is a decreasing function.  If we replace the kernel function $K(\cdot)$ by the von Mises-Fisher kernel, the above expression becomes
\[
\hat{\boldsymbol{x}}^{(s+1)} =  \frac{\sum_{i=1}^n \boldsymbol{x}_i \exp\biggl(\frac{\hat{\boldsymbol{x}}^{(s)\top} \boldsymbol{x}_i-1}{h^2} \biggr)}{\Big\vert\Big\vert \sum_{i=1}^n \boldsymbol{x}_i \exp\biggl( \frac{\hat{\boldsymbol{x}}^{(s)\top} \boldsymbol{x}_i -1}{h^2} \biggr)\Big\vert\Big\vert_2}.
\]

Straightforward calculations allowed us to extend the proposal by \cite{Yang2014} to varying $h_i$, that is, for adaptive kernel density estimation on the unit sphere.

\subsection{Post-processing}
\label{par:signif_vmf}
As mentioned in Section~\ref{high-energy-astrophysics}, the incoming photons were either emitted from a high-energy source or are part of the diffuse $\gamma$-ray background which spreads over the entire area observed by the telescope.  The directional kernel density estimator \eqref{vMF_directional_kernel} tries and reconstructs the corresponding mixture distribution.  Hence, the small peaks which emerge as modes may identify true sources, but they may equally well represent a false signal generated by the irregularly shaped background radiation.  To separate the true signal emitted by a source from the background, we developed a post-processing procedure that combines the findings of two parallel quests.  One establishes the significance of a candidate mode using a suitable statistical test.  The second skims off the photons emitted by the $\gamma$-ray background using a suitable classifier build on previous knowledge provided by the Fermi LAT Collaboration.  By super-imposing the findings from these two quests, we identify candidate sources which are both, statistically significant and qualified as such according to a set of relevant features.  Furthermore, we are now able to distinguish photons emitted by a candidate source from those pertaining to the background radiation.

\subsubsection{Statistical significance}
\label{statistical-significance}
Mathematically, we can verify whether a function reaches a local maximum by checking whether all eigenvalues of the Hessian matrix evaluated at the candidate mode are negative.  Statistically, developing a suitable test to verify the existence of a mode and deriving its null distribution using eigenvalues is tricky, as these are not continuously differentiable functions of the Hessian.  This invalidates resampling-based methods such as the bootstrap and asymptotic expansion by the delta method, which we may use to reconstruct the finite-sample null distribution of the test statistic.  \cite{Genovese2016} hence suggest to use data splitting to separate the process of finding candidate modes from the process of hypothesis testing.  They furthermore propose to base inference on confidence intervals, rather than on $p$-values.  The potential modes are hence estimated on the first half of the data, while the second half is used to construct asymptotically valid bootstrap confidence intervals for the eigenvalues of the Hessian matrix, which can be used for hypothesis testing. 

The extension of this idea to directional data requires some care, as working on the unit sphere sets some constraints.  To calculate the Hessian matrix $\mathcal{H} \hat f_h(\boldsymbol{x})$, we first need the total gradient 
\[
\nabla \hat f_h(\boldsymbol{x}) = \frac{C_2(h^{-2})}{n}\sum_{i=1}^n \frac{\boldsymbol{x}_i}{h^2}\exp\Bigg( \frac{\boldsymbol{x}^\top \boldsymbol{x}_i -1 }{h^2}\Bigg),
\]
where $\nabla$ represents suitable differentiation.  The Hessian matrix hence is
\begin{equation}
\label{hess_riemann}
\nonumber
\begin{split}
\mathcal{H}\hat f_h(\boldsymbol{x}) & = (\text{\bf I}_3 - \boldsymbol{x}\boldsymbol{x}^\top) \Big( \nabla \nabla \hat f_h(\boldsymbol{x}) - \nabla \hat f_h(\boldsymbol{x})^\top \boldsymbol{x} \text{\bf I}_3 \Big)(\text{\bf I}_3 - \boldsymbol{x}\boldsymbol{x}^\top) \\
& = (\text{\bf I}_3 - \boldsymbol{x}\boldsymbol{x}^\top) \Bigg[\frac{C_{2}(h^{-2})}{n}\sum_{i=1}^n \frac{\boldsymbol{x}_i\boldsymbol{x}_i^\top}{h^4}\exp\Bigg( \frac{\boldsymbol{x}^\top \boldsymbol{x}_i -1 }{h^2}\Bigg) + \\
& -  \frac{C_{2}(h^{-2})}{n}\sum_{i=1}^n \frac{\boldsymbol{x}^\top \boldsymbol{x}_i \text{\bf I}_3}{h^2}\exp\Bigg( \frac{\boldsymbol{x}^\top \boldsymbol{x}_i -1 }{h^2}\Bigg)\Bigg] (\text{\bf I}_3 - \boldsymbol{x}\boldsymbol{x}^\top).
\end{split}
\end{equation}
Likewise, we may obtain the Hessian matrix associated with an adaptive kernel density estimator with variable bandwidth $h_i$.  The tricky part is that the eigenvalue of $\mathcal{H} \hat f_h(\boldsymbol{\mu})$, when $\hat f_h(\boldsymbol{x})$ is evaluated at $\boldsymbol{\mu}$, is always zero, whether $\boldsymbol{\mu}$ corresponds to a true source or not.  This entails that inference has to be based on the remaining two eigenvalues.  We hence construct an $1-\alpha$ level confidence interval for the largest non null eigenvalue using bootstrap resampling.  The candidate mode is validated if the interval includes only negative values.  

A second possibility is to reparametrize the von Mises-Fisher kernel using polar coordinates
\[
f_{vMF}(\theta, \phi) = \frac{\kappa}{4\pi\kappa} \exp\Big[\kappa \cos \theta \cos\eta + k\sin \theta \sin\eta\cos(\phi-\zeta) \Big]\sin\theta,
\]
where, as in \cite{Mardia2000}, $\boldsymbol{x} = (\cos\theta,\, \sin \theta \cos\phi,\, \sin\theta\sin\phi)^\top$ and $\boldsymbol{\mu} = (\cos\eta,\, \sin\eta\cos\zeta,\, \sin\eta\sin\zeta)^\top$.  This workaround allows us to directly apply the results by \cite{Genovese2016}.

\subsubsection{Feature selection}
\label{feature-selection}
A further possibility to skim off the photons emitted by extra-galactic sources from those which originate from the diffuse background is to build a suitable classification rule which integrates additional information on the photons provided by the Fermi LAT Collaboration and/or features that can be extracted at the various steps of the mean-shift algorithm.  These include the energy content of the photons ({\it photon\_energy}) and their incoming direction ({\it longitude}, {\it latitude}), the number of photons assigned to a mode ({\it n\_photons}), the density estimates for the signal and the background model ({\it density}, {\it density\_difference}) and various types of distances between the photons and their mode ({\it intra\_cluster\_distance}, {\it total\_distance}, {\it first\_step\_length}).  We hence suggest to train and test a tree-based classifier on a suitable area of the sky.  The final classifier will then be pruned so as to assign any cluster with a single photon to the background.  Section~\ref{performance-metrics} reports the performance metrics of our classification rule when applied to a portion of the Northern sky.

\section{Application to Fermi LAT data}

\subsection{Benchmarking}
\label{benchmarking}

\begin{SCtable}[0.9][t]
\begin{tabular}{lccc}
\toprule
$h$ & ARI & $\bar d(s, \hat s)$ & $n_{s}$\\ 
\hline
$h_{i,SE}$ & 0.9976 & 0.0004 & 86\\ 
$h_{THUMB}$ & 0.6841 & 0.0079 & 10\\
$h^A_{i,THUMB}$ & 0.6805 & 0.0139 & 18\\
$h^S_{i,THUMB}$ & 0.8524 & 0.0063 & 25\\
$h_{LCV}$ & 0.9777 & 0.0092 & 142\\ 
$h^A_{i,LCV}$ & 0.9777 & 0.0092 & 142\\ 
$h^S_{i,LCV}$ & 0.9777 & 0.0092 & 142\\ 
\hline
\end{tabular}
\caption{Performance metrics for different choices of the bandwidth $h$ of the von Mises-Fisher kernel density estimator applied to the sky region plotted in Figure~\ref{fig:4}: ARI = adjusted Rand index; $\bar d (s,\hat s)$ = median angular distance (in degrees) between the directions of true sources ($s$) and candidate sources ($\hat s$) identified by the algorithm; $n_s$ = number of identified sources.  The number of true sources is $68$.}
\label{tab:1}
\end{SCtable}

\subsubsection{Optimal bandwidth}
\label{optimal-bandwidth}
Table~\ref{tab:1} compares the different proposals for bandwidth selection listed in Sections~\ref{data-based-methods} and \ref{using-scientific-input} using three performance metrics, that is, the adjusted Rand index (ARI), the median angular distance (in degrees) between the directions of true sources and candidate sources, $\bar d(s,\hat s)$, and the number $n_s$ of identified sources.  These metrics were obtained by benchmarking our algorithm on a simulated sample of 2.335 photons emitted by the 68 sources present in the validation region $(l,b) \in [0^\circ,60^\circ] \times [10^\circ,60^\circ]$ shown in Figure~\ref{fig:4}.  The three proposals based on the rule of thumb oversmooth the true photon density, leading to rather low ARI values.  Likelihood cross validation, on the other hand, tends to over adapt the true density yielding too many candidate sources: 142 in place of the 68 present.  The best partition of the selected sky region is obtained when using the variable bandwidth $h_{i,SE}$, that is, the scale factor of the LAT's point spread function.  
Further support to this choice is provided by Table~\ref{tab:2}, which contrasts the selected optimal bandwidths (Columns~3--7) with the true photon scattering, as measured by its standard deviation (Column 2), for 5 selected sources of varying size, that is, which emit from a minimum of $n_s=7$ photons up to a mximum of $n_s=151$ photons.  Again, $h_{i,SE}$ is the best performing choice.
 
\begin{table}[t]
\centering
\renewcommand\arraystretch{1.2}
\begin{tabular}{lcccccc}
\toprule
Source &sd & $\bar h_{i, SE}$ & $\bar h^A_{i, LCV}$ & $\bar h^A_{i,THUMB}$ & $\bar h^S_{i, LCV}$ & $\bar h^S_{i,THUMB}$\\ 
\hline
$n_s=7$ & 0.0019 & 0.0017 & $3.2958 \cdot 10^{-06}$ & 0.1053 & $2.8623\cdot 10^{-07}$ & 0.0611 \\
$n_s=19$ & 0.0048 & 0.0028  & $3.2225\cdot 10^{-06}$ & 0.0221& $2.7986\cdot 10^{-07}$ & 0.0128 \\
$n_s=31$ & 0.0042 & 0.0027  & $2.8184 \cdot10^{-06}$ & 0.0501 & $2.4477\cdot 10^{-07}$ & 0.0290\\
$n_s=79$ & 0.0030 & 0.0028  & $2.1721\cdot 10^{-06}$ & 0.0314 & $1.8864\cdot 10^{-07}$ & 0.0182\\
$n_s=151$ & 0.0068 & 0.0027  & $2.0684 \cdot 10^{-06}$ & 0.0215 & $1.7963 \cdot 10^{-07}$ & 0.0125\\
\hline
\end{tabular}
\caption{Standard deviation (Column~2) of photon scattering for 5 selected sources of varying size (Column~1) and average bandwidths computed using the scale factor of the PSF (Column~3) or selected by Abramson's or Silverman's rules (Columns~4--7).}
\label{tab:2}
\end{table}

\subsubsection{Performance metrics}
\label{performance-metrics}

We bemchmarked our tree-based classifier on a sample of 35,365 simulated photon emissions in the sky region $(l,b) \in [100^\circ,150^\circ] \times [0^\circ,90^\circ]$.  This area covers the entire Northern sky to account for the rater prominent variability of the diffuse $\gamma$-ray background as we move away from the Galactic plane.  The classifier was estimated on the first 2/3 of the sample, for a total of 24,573 photons, and tested on the remaining 11,062 photons.  In both sets, about 85\% of the photons were emitted from the background.  The final classifier was pruned so as to assign any cluster with a single photon to the background.  The classifier was hence benchmarked on the sky region shown in Figure~\ref{fig:4}, where it selected a total of $n_s=86$ sources.  The average sensibility, computed on the candidate sources identified by the classifier, was 90,5\%, while the average specificity was 99,5\%.  The adjusted Rand index (ARI) is 0.9752 and the median angular distance between the true sources and the identified ones is 0.0005 degrees.

\subsection{Case-study}
\label{case-study}
\begin{figure}[t]
\centering
\quad
\includegraphics[height=0.22\textheight]{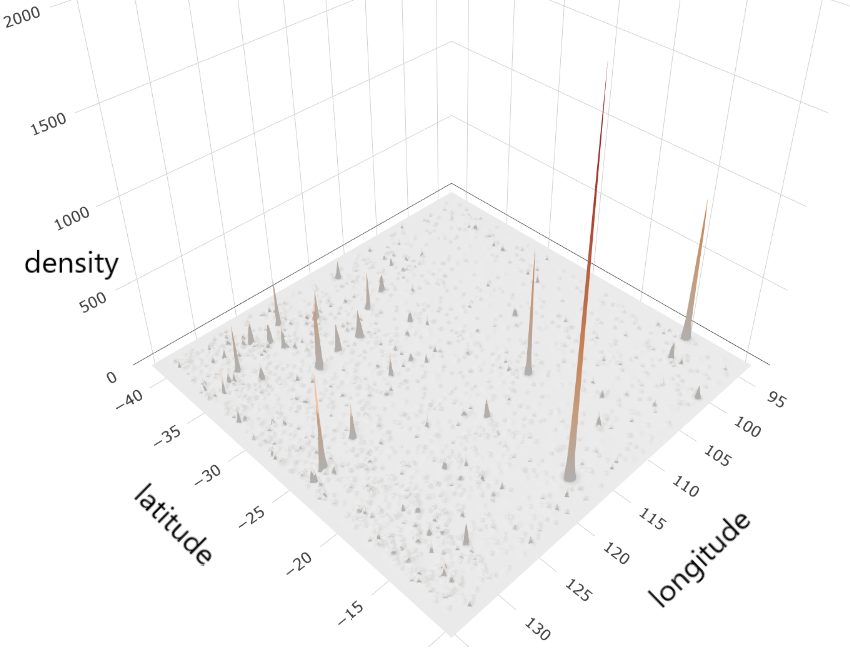} 
\quad 
\includegraphics[height=0.16\textheight]{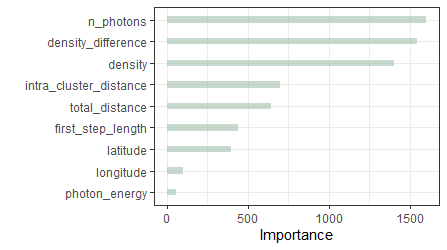} 
\caption[t]{Left: Kernel density estimate using a von Mises-Fisher kernel for the 3,849 $\gamma$-ray photon counts accumulated by the LAT in a 5-year period in the region of the Southern sky identified by $(l,b)\in[95^\circ,135^\circ] \times [-40^\circ,-10^\circ]$.  About 26\% of the photons were
emitted by the 44 sources present in the area.  Right: Feature importance plot for the tree-based photon classifier used to discriminate between source and diffuse $\gamma$-ray background emission.}
\label{fig:5}       
\end{figure}

The yellow region in Figure~\ref{fig:1} shows a portion of the Southern sky of size $(l,b)\in[95^\circ,135^\circ]\times[-40^\circ,-10^\circ]$ for which the LAT accumulated 3,849 photon counts over a five-year period of observation.\footnote{\url{https://fermi.gsfc.nasa.gov/ssc/data/access/}}  Of these, about 26\% were emitted by the 44 sources present in the area, while the remaining 74\% originated from the diffuse $\gamma$-ray background.  The left panel of Figure~\ref{fig:5} plots the estimated kernel density \eqref{vMF_directional_kernel} using a von Mises-Fisher kernel.  Here, the bandwidth parameter $h$ was set according to scientific input, as described in Section~\ref{using-scientific-input}.  This choice revealed to be the most performing one in terms of adjusted Rand index (ARI), median angular distance, $\bar d(s,\hat s)$, between the true source direction and the reconstructed one and number of identified sources ($n_s$).  In all, the mean-shift algorithm identified 876 modes.   To further refine the list of candidate sources we proceeded in two steps as outlined in Section~\ref{par:signif_vmf}.  

\begin{figure}[t]
\centering
\begin{minipage}[]{.65\textwidth}
\flushleft
\includegraphics[width=0.95\textwidth]{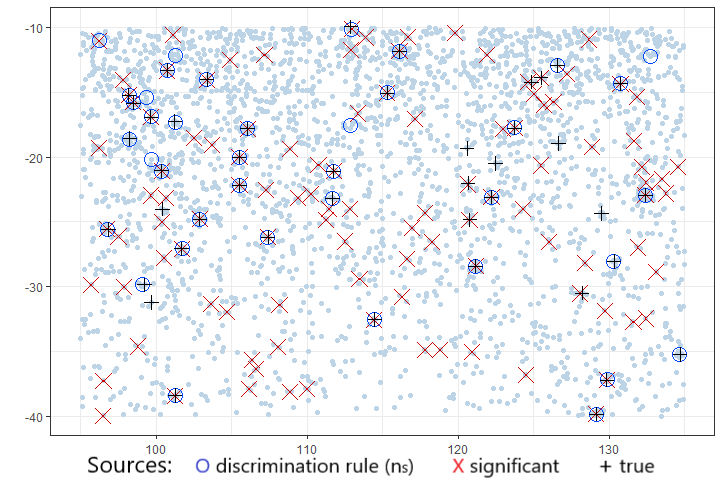}
\label{fig:3} 
\end{minipage}
\begin{minipage}[]{.05\textwidth}
\flushleft
{\small
\begin{tabular}{cc}
\toprule
\multicolumn{2}{c}{Results} \\
\hline
ARI & 0.961 \\ 
$\bar d(s, \hat s)$ & 0.001 \\
$n_{s}$ & 39\\
True sources & 44 \\
\hline
\end{tabular}}
\end{minipage}\hspace*{\fill}
\caption{Left: Fermi-LAT $\gamma$-ray photon count map (in Galactic coordinates) for the analysed \mbox{5-year} observation period with superimposed the true (black crosses) and candidate sources.  A red cross pinpoints a candidate source which is statistically significant at the 5\% level, while a blue circle identifies a candidate source on the basis of its features.  Right: Performance measures of the tree-based classifier.}
\label{fig:6}
\end{figure}

A tree-based classifier to discriminate between source and background photons was trained on the 6,814 photon counts highlighted in green in Figure~\ref{fig:1}.  The importance of the selected predictor variables is shown in the right panel of Figure~\ref{fig:5}.  The most discriminating features are the number of photons assigned to a cluster ({\it n\_photons}), the difference between the two photon densities for, respectively, the all sky and background counts only ({\it density\_differences}), and the density observed for each photon ({\it density}).  This reduces the original 876 modes to 39 candidate sources, which are shown as blue circles in the left panel of Figure~\ref{fig:6}.  The table on the right reports the performance of our classifier in terms of ARI and median angular distance $\bar d(s, \hat s)$.  The true positive rate for single photon classification is 98.5\% rate, while the percentage of false positives is 22.9\%.  Indeed, the five missed sources are the less photon emitting ones. 

In parallel, we tested all the 555 clusters which contain two or more photons at a significance level of 5\% as outlined in Section~\ref{statistical-significance} and applying Bonferroni's correction.  This skimmed off 448 modes, for a total of 107 remaining candidate sources, shown in the left panel of Figure~\ref{fig:3} as red crosses.  Here, the true positive rate for single photon classification is 85.0\% and the false positive rate is 11.2\%. 

By super-imposing these two findings, we obtain in all 27 sources which are both, statistically significant and qualified as such by the non-parametric classifier.  The global true positive rate for single photon classification is 94.6\% while the false positive rate is 14.1\%.

\section{Concluding remarks}
\label{conclusions}

Astronomical data typically come in the form of {\it big data}, whose volumes have increased over the past years from gigabytes into terabytes and petabytes.  However, the widely used model-based approach to multivariate classification, which involves maximizing the likelihood of the mixture model using e.g. the expectation maximization (EM) algorithm or Markov chain Monte Carlo (McMC) simulation, is computationally impractical for today’s enormous databases.  Suitable machine-learning techniques, that apply to such volumes of data, have recently made their way into the general knowledge basis of the astrophysics community.  Yet, they miss the flexibility and adaptability which is required if we want to take account of further pieces of available information, such as the energy content and quality of the detected events and/or temporal aspects.  Indeed, this may allow us to more efficiently determine the physical origin of the signals and to discover rare and/or very faint objects, leading to major discoveries in astrophysics.

Our proposal represents a fast and scalable computational tool to efficiently and effectively extract knowledge from such large databases.  As our aim is to analyze whole sky maps in one go, we are currently fine-tuning our algorithm by including a {\it consensus clustering} step.  This will allow us to aggregate results from multiple runs, while guaranteeing more stable and robust results \citep{Monti2003,VegaPons2011}.  More precisely, borrowing from \cite{NordhaugMyhre2018}, we form a {\it clustering ensemble} consisting of separate and bootstrapped runs of the mean-shift algorithm on a given number of overlapping regions of the sky, as shown in Figure~\ref{fig:3}.  The size and location of these regions varies on a random basis.  The final modes are identified by selecting the cluster configuration  which was observed most of the times.  This way of proceeding guarantees robustness with respect to the choice of the smoothing parameter $h$, while at the same time allowing us to work with tremendous amount of data.

\begin{figure}[t]
\centering
\includegraphics[height=0.22\textheight]{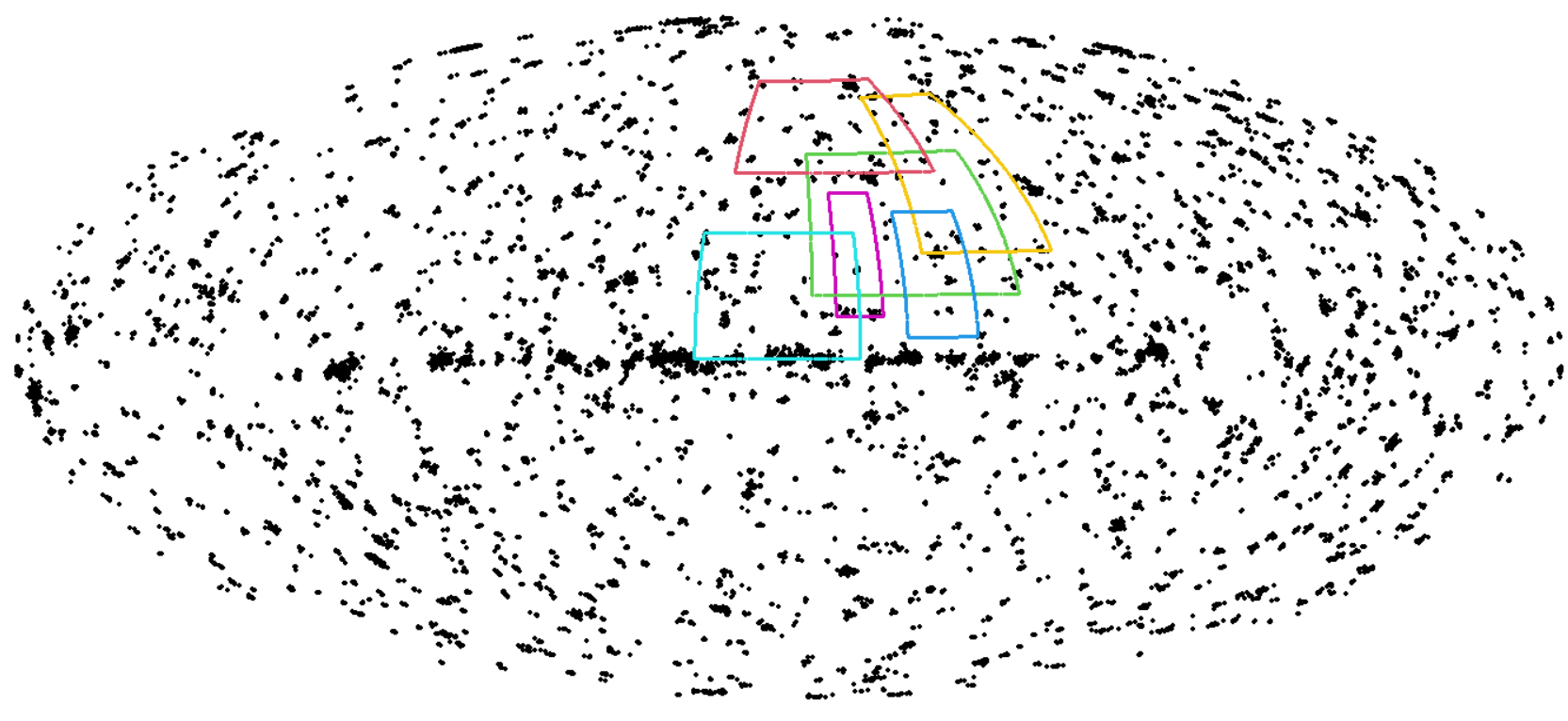}
\caption{Selection of overlapping regions for {\it consensus} clustering.}
\label{fig:3}       
\end{figure}

\section*{Competing interests and funding}
The authors have no relevant financial or non-financial interests to disclose.

\bibliography{sn-bibliography}

\begin{thebibliography}{}
\providecommand{\doi}[1]{\url{https://doi.org/#1}}
\bibcommenthead

\bibitem[\protect\citeauthoryear{Abramson}{Abramson}{1982}]{Abramson0a}
Abramson, I.S. 1982.
\newblock {Arbitrariness of the pilot estimator in adaptive kernel methods}.
\newblock {\em Journal of Multivariate Analysis\/}~{\em 12\/}(4): 562--567 .

\bibitem[\protect\citeauthoryear{Acero, Ackermann, Ajello, Albert, Baldini,
  Ballet, and Barbiellini}{Acero et~al.}{2016}]{Acero_2016}
Acero, F., M.~Ackermann, M.~Ajello, A.~Albert, L.~Baldini, J.~Ballet, and
  G.~Barbiellini. 2016.
\newblock Development of the model of galactic interstellar emission for
  standard point source analysis of {Fermi Large Area Telescope} data.
\newblock {\em The Astrophysical Journal Supplement Series\/}~{\em 223\/}(2):
  26 .

\bibitem[\protect\citeauthoryear{Ackermann, Ajello, and Allafort}{Ackermann
  et~al.}{2013}]{Ackermann2013}
Ackermann, M., M.~Ajello, and Allafort. 2013.
\newblock Determination of the point-spread function for the {Fermi} large area
  telescope from on-orbit data and limits on pair halos of active galactic
  nuclei.
\newblock {\em Astrophysical Journal\/}~{\em 765\/}(1) .

\bibitem[\protect\citeauthoryear{Bai, Rao, and Zhao}{Bai
  et~al.}{1988}]{Bai1988}
Bai, Z.D., C.R. Rao, and L.C. Zhao. 1988.
\newblock {Kernel estimators of density function of directional data}.
\newblock {\em Journal of Multivariate Analysis\/}~{\em 27\/}(1): 24--39 .

\bibitem[\protect\citeauthoryear{Banerjee, Dhillon, Ghosh, and Sra}{Banerjee
  et~al.}{2006}]{Banerjee2006}
Banerjee, A., I.S. Dhillon, J.~Ghosh, and S.~Sra. 2006.
\newblock {Clustering on the Unit Hypersphere using von Mises-Fisher
  Distributions}.
\newblock {\em ournal of Machine Learning Research\/}~6: 1345--1382 .

\bibitem[\protect\citeauthoryear{Carmichael, George, and Julius}{Carmichael
  et~al.}{1968}]{Carmichael1968}
Carmichael, J.W., J.A. George, and R.S. Julius. 1968.
\newblock {Finding natural clusters}.
\newblock {\em Systematic Biology\/}~{\em 17\/}(2): 144--150 .

\bibitem[\protect\citeauthoryear{Chac{\'{o}}n}{Chac{\'{o}}n}{2015}]{Chacon2015}
Chac{\'{o}}n, J.E. 2015.
\newblock {A population background for nonparametric density-based clustering}.
\newblock {\em Statistical Science\/}~{\em 30\/}(4): 518--532 .

\bibitem[\protect\citeauthoryear{Costantin, Menardi, Brazzale, Bastieri, and
  Fan}{Costantin et~al.}{2020}]{Costantin2020primo}
Costantin, D., G.~Menardi, A.R. Brazzale, D.~Bastieri, and J.H. Fan. 2020.
\newblock {A novel approach for pre-filtering event sources using the von
  Mises–Fisher distribution}.
\newblock {\em Astrophysics and Space Science\/}~365: 53 .

\bibitem[\protect\citeauthoryear{Costantin, Sottosanti, Brazzale, Bastieri, and
  Fan}{Costantin et~al.}{2020}]{Costantin2020due}
Costantin, D., A.~Sottosanti, A.R. Brazzale, D.~Bastieri, and J.H. Fan. 2020.
\newblock {Bayesian mixture modelling of the high-energy photon counts
  collected by the Fermi Large Area Telescope}.
\newblock {\em Statistical Modelling\/}~{\em 22\/}(3): 1--37 .

\bibitem[\protect\citeauthoryear{Fraley and Raftery}{Fraley and
  Raftery}{1998}]{FraleyRaftery1998}
Fraley, C. and A.E. Raftery. 1998.
\newblock How many clusters? which clustering method? answers via model-based
  cluster analysis.
\newblock {\em The Computer Journal\/}~{\em 41\/}(8): 578--588 .

\bibitem[\protect\citeauthoryear{Fraley and Raftery}{Fraley and
  Raftery}{2002}]{FraleyRaftery2002}
Fraley, C. and A.E. Raftery. 2002.
\newblock Model-based clustering, discriminant analysis, and density
  estimation.
\newblock {\em Journal of the American Statistical Association\/}~{\em
  97\/}(458): 611--631 .

\bibitem[\protect\citeauthoryear{Fukunaga and Hostetler}{Fukunaga and
  Hostetler}{1975}]{Fukunaga1975}
Fukunaga, K. and L.D. Hostetler. 1975.
\newblock {The Estimation of the Gradient of a Density Function, with
  Applications in Pattern Recognition}.
\newblock {\em IEEE Transactions on Information Theory\/}~{\em 21\/}(1): 32--40
  .

\bibitem[\protect\citeauthoryear{Garc{\'{i}}a-Portugu{\'{e}}s}{Garc{\'{i}}a-Portugu{\'{e}}s}{2013}]{Garcia-Portugues2013}
Garc{\'{i}}a-Portugu{\'{e}}s, E. 2013.
\newblock {Exact risk improvement of bandwidth selectors for kernel density
  estimation with directional data}.
\newblock {\em Electronic Journal of Statistics\/}~{\em 7\/}(1): 1655--1685 .

\bibitem[\protect\citeauthoryear{Genovese, Perone-Pacifico, Verdinelli, and
  Wasserman}{Genovese et~al.}{2016}]{Genovese2016}
Genovese, C.R.., M.~Perone-Pacifico, I.~Verdinelli, and L.~Wasserman. 2016.
\newblock {Non-parametric inference for density modes}.
\newblock {\em Journal of the Royal Statistical Society\/}~{\em 78\/}(1):
  99--126 .

\bibitem[\protect\citeauthoryear{Hall, Watson, and Cabrera}{Hall
  et~al.}{1987}]{10.1093/biomet/74.4.751}
Hall, P., G.S. Watson, and J.~Cabrera. 1987.
\newblock {Kernel density estimation with spherical data}.
\newblock {\em Biometrika\/}~{\em 74\/}(4): 751--762 .

\bibitem[\protect\citeauthoryear{Hartigan}{Hartigan}{1975}]{Hartigan75}
Hartigan, J.A. 1975.
\newblock {\em Clustering Algorithms}.
\newblock John Wiley \& Sons.

\bibitem[\protect\citeauthoryear{Hennig, Meila, Murtagh, and Rocci}{Hennig
  et~al.}{2015}]{Hennig-etal-2015}
Hennig, C., M.~Meila, F.~Murtagh, and R.~Rocci. 2015.
\newblock {\em Handbook of Cluster Analysis}.
\newblock Chapman and Hall/CRC. CRC Press.

\bibitem[\protect\citeauthoryear{Hobson, Jaffe, Liddle, Mukherjee, and
  Parkinson}{Hobson et~al.}{2009}]{Hobson-etal-2009}
Hobson, M., A.~Jaffe, A.~Liddle, P.~Mukherjee, and D.~Parkinson eds. 2009.
\newblock {\em Bayesian Methods in Cosmology}.
\newblock Cambridge University Press.

\bibitem[\protect\citeauthoryear{Izenman}{Izenman}{1991}]{Izenman91}
Izenman, A.J. 1991.
\newblock Review papers: Recent developments in nonparametric density
  estimation.
\newblock {\em Journal of the American Statistical Association\/}~86: 205--224
  .

\bibitem[\protect\citeauthoryear{Jones, Kashyap, and {Van Dyk}}{Jones
  et~al.}{2015}]{Jones2015}
Jones, D.E., V.L. Kashyap, and D.A. {Van Dyk}. 2015.
\newblock Disentangling overlapping astronomical sources using spatial and
  spectral information.
\newblock {\em Astrophysical Journal\/}~{\em 808\/}(2) .

\bibitem[\protect\citeauthoryear{Klemelä}{Klemelä}{2000}]{KLEMELA200018}
Klemelä, J. 2000.
\newblock Estimation of densities and derivatives of densities with directional
  data.
\newblock {\em Journal of Multivariate Analysis\/}~{\em 73\/}(1): 18--40 .

\bibitem[\protect\citeauthoryear{Mardia and Jupp}{Mardia and
  Jupp}{2000}]{Mardia2000}
Mardia, K.V. and P.E. Jupp. 2000.
\newblock {\em Directional Statistics}.
\newblock John Wiley \& Sons.

\bibitem[\protect\citeauthoryear{Mattox, Bertsch, Chiang, and Dingus}{Mattox
  et~al.}{1996}]{MattoxJ}
Mattox, J.R., D.L. Bertsch, J.~Chiang, and B.L. Dingus. 1996.
\newblock {The likelihood analysis of EGRET data}.
\newblock {\em The Astrophysical Journal\/}~148: 148--162 .

\bibitem[\protect\citeauthoryear{Menardi}{Menardi}{2016}]{Menardi2016}
Menardi, G. 2016.
\newblock A review on modal clustering.
\newblock {\em International Statistical Review\/}~{\em 84\/}(3): 413--433 .

\bibitem[\protect\citeauthoryear{Meyer, van Dyk, Kashyap, Campos, Jones,
  Siemiginowska, and Zezas}{Meyer et~al.}{2021}]{Meyer2021}
Meyer, A.D., D.A. van Dyk, V.L. Kashyap, L.F. Campos, D.E. Jones,
  A.~Siemiginowska, and A.~Zezas. 2021, 06.
\newblock {eBASCS: Disentangling overlapping astronomical sources II, using
  spatial, spectral, and temporal information}.
\newblock {\em Monthly Notices of the Royal Astronomical Society\/}~{\em
  506\/}(4): 6160--6180 .

\bibitem[\protect\citeauthoryear{Milnor, Spivak, and Wells}{Milnor
  et~al.}{1969}]{Milnor1969}
Milnor, J., M.~Spivak, and R.~Wells. 1969.
\newblock {\em Morse Theory. (AM-51), Volume 51}.
\newblock Princeton University Press.

\bibitem[\protect\citeauthoryear{Monti, Tamayo, Mesirov, and Golub}{Monti
  et~al.}{2003}]{Monti2003}
Monti, S., P.~Tamayo, J.~Mesirov, and T.~Golub. 2003.
\newblock {Consensus clustering: a resampling-based method for class discovery
  and visualization of gene expression microarray data}.
\newblock {\em Machine Learning\/}~52: 91--118 .

\bibitem[\protect\citeauthoryear{Montin, Brazzale, and Menardi}{Montin
  et~al.}{2022}]{Montin2022}
Montin, A., A.R. Brazzale, and G.~Menardi 2022.
\newblock Locating $\gamma$-ray sources on the celestial sphere via modal
  clustering.
\newblock In A.~Balzanella, M.~Bini, C.~Cavicchia, and R.~Verde (Eds.), {\em
  Book of the Short Papers -- SIS 2022}, pp.\  1582--1587.

\bibitem[\protect\citeauthoryear{{Nordhaug Myhre}, {{\O}yvind Mikalsen},
  L{\o}kse, and Jenssen}{{Nordhaug Myhre} et~al.}{2018}]{NordhaugMyhre2018}
{Nordhaug Myhre}, J., K.~{{\O}yvind Mikalsen}, S.~L{\o}kse, and R.~Jenssen.
  2018.
\newblock {Robust clustering using a kNN mode seeking ensemble}.
\newblock {\em Pattern Recognition\/}~76: 491--505 .

\bibitem[\protect\citeauthoryear{Silverman}{Silverman}{1986}]{silverman2018density}
Silverman, B.W. 1986.
\newblock {Density Estimation for Statistics and Data Analysis}.
\newblock {\em Statistics and Applied Probability\/} .

\bibitem[\protect\citeauthoryear{Sottosanti, Bernardi, Brazzale,
  Geringer-Sameth, Stenning, Trotta, and van Dyk}{Sottosanti
  et~al.}{2021}]{Sottosanti2021}
Sottosanti, A., M.~Bernardi, A.R. Brazzale, A.~Geringer-Sameth, D.C. Stenning,
  R.~Trotta, and D.A. van Dyk. 2021.
\newblock {Identification of high-energy astrophysical point sources via
  hierarchical Bayesian nonparametric clustering}.
\newblock {\em arXiv preprint arXiv:2104.11492\/} .

\bibitem[\protect\citeauthoryear{van Dyk, Connors, Kashyap, and
  Siemiginowska}{van Dyk et~al.}{2001}]{van_Dyk_2001}
van Dyk, D.A., A.~Connors, V.L. Kashyap, and A.~Siemiginowska. 2001.
\newblock Analysis of energy spectra with low photon counts via bayesian
  posterior simulation.
\newblock {\em The Astrophysical Journal\/}~{\em 548\/}(1): 224--243 .

\bibitem[\protect\citeauthoryear{Vega-Pons and Ruiz-Shulcloper}{Vega-Pons and
  Ruiz-Shulcloper}{2011}]{VegaPons2011}
Vega-Pons, S. and J.~Ruiz-Shulcloper. 2011.
\newblock {A survey of clustering ensemble algorithms}.
\newblock {\em International Journal of Pattern Recognition and Artificial
  Intelligence\/}~{\em 25\/}(3): 337--372 .

\bibitem[\protect\citeauthoryear{Yang, Chang-Chien, and Kuo}{Yang
  et~al.}{2014}]{Yang2014}
Yang, M.S., S.J. Chang-Chien, and H.C. Kuo 2014.
\newblock On mean shift clustering for directional data on a hypersphere.
\newblock In L.~Rutkowski, M.~Korytkowski, R.~Scherer, R.~Tadeusiewicz, L.~A.
  Zadeh, and J.~M. Zurada (Eds.), {\em Artificial Intelligence and Soft
  Computing}, Cham, pp.\  809--818. Springer International Publishing.

\end{thebibliography}

\end{document}